\documentclass[12pt,a4paper]{revtex4}
\usepackage{graphicx}
\usepackage{txfonts}
\usepackage{amssymb}
\usepackage{natbib}

\begin{document}
\title{The constructive role of diversity on the global response of coupled neuron systems}
\author{Toni P\'erez$^{1}$, Claudio R. Mirasso$^2$, Ra\'ul Toral$^2$ and James
D. Gunton$^1$}
\affiliation{$^1$Department of Physics, Lehigh University, Bethlehem,
Pennsylvania
18015, USA\\
$^2$IFISC (Instituto de F{\'\i}sica Interdisciplinar y Sistemas
Complejos), UIB-CSIC. Campus UIB, 07122 Palma de Mallorca,Spain}
\begin{abstract}

We study the effect that the heterogeneity present among the elements of an
ensemble of coupled
excitable neurons have on the collective response of the system to an external
signal. We have considered
two different interaction scenarios, one in which the neurons are diffusively
coupled and another in which
the neurons interact via pulse-like signals. We found that the
type of interaction between the neurons has a crucial role in determining the
response of the system to
the external modulation. We develop a mean-field theory based on an order
parameter expansion that
quantitatively reproduces the numerical results in the case of diffusive
coupling.
\end{abstract}

\maketitle

\section{Introduction}
\label{introduction}
Synchronized behavior arising among the constituents of an ensemble is common in
nature. Examples include the synchronized flashing of fireflies, the
blossoming of flowers, cardiac cells giving rise to the pacemaker role of the
sinoatrial node of the heart and the electrical pulses of neurons.  This global
behavior can  originate from a common response to an external stimulus or might
appear in autonomous, non-forced, systems.  The theoretical basis for the
understanding of synchronization in non-forced systems was put forward by
Winfree
(\cite{Winfree67}) who showed that the interaction --i.e. coupling-- between
the constituents is  an essential ingredient for the existence of a synchronized
output. The seminal work on coupled oscillators by Kuramoto
(\cite{Kuramoto1975})
offered a model case whose solution confirms the basic hypothesis of Winfree:
while interaction helps towards the achievement of a common behavior, a perfect
order can be achieved only in the absence of diversity --heterogeneity-- among
the components of an ensemble. In the Kuramoto model, diversity manifests when
the oscillators have different natural frequencies, those they display when
uncoupled from each other. While this result holds for systems that can be
described by coupled oscillators, recent results indicate that in other cases
diversity among the constituents might actually have a positive role in the
setting of a resonant behavior with an external signal. This was first
demonstrated in
(\cite{Tessone2006}) and has been since extended to many other systems
(\cite{Gosak2009,Ullner2009,Zanette2009,Tessone2009,Postnova09,Chen09,Wu09,
Zanette2009,
Ullner2009,Perc08,Acebron07}). In the case of non-forced excitable systems, a
unifying treatment of the role of noise and diversity has been developed in
\cite{TSTP:2007}.

Many biological systems, including neurons, display excitability as a response
to an
external perturbation. Excitability is
characterized by the existence of a threshold, a largely independent
response to a suprathreshold input and a refractory time. It is well established
that the dynamical features of neurons can be described by excitable models: when a neuron is
perturbed by a single impulse, the neuron can generate a single spike, and when perturbed by a
continuum signal, a train of spikes with a characteristic firing frequency can appear instead.
Although the creation and propagation of electrical signals has been thoroughly studied by physiologists for at least
a century, the most important landmark in this field is due to Hodgkin and Huxley
(\cite{HH1952d}), who developed the first quantitative model to describe the
evolution of the membrane potential in the squid giant axon. Because of the
central importance of cellular electrical activity in biological systems and
because this model forms the basis for the study of excitable systems, it
remains to this date an important model for analysis.  Subsequently a simplified
version of this model, known as the FitzHugh-Nagumo (FHN) model
(\cite{FHN1961}), that captures many of the qualitative features of the
Hodgkin-Huxley model was developed. The FHN model has two variables, one fast
and one slow. The fast variable represents the evolution of the membrane potential and it is
known as the excitation variable. The slow variable accounts for the $K^{+}$
ionic current and is known as the recovery variable.  One virtue of this model is that it can be studied
using phase-plane methods, because it is a two variable system.  For instance,
the fast variable has a cubic nullcline while the slow variable has a linear
nullcline and the study of both nullclines and their intersections allow to
determine the different dynamical regimes of the model. The FHN model
can then be extended by coupling the individual elements, and provides an
interesting model, for example, of coupled neuron populations.

In the initial studies of the dynamics of the FHN and similar models,
the individual units (e.g. neurons) were treated as identical. However, it is
evident that real populations of neurons
display a large degree of variability, both in morphology and dynamical
activity; that is, there is a diversity in the population of these biological units. Then, it is natural
to ask what role diversity plays in the global dynamical behavior of these systems and a lot
of activity has been developed along these lines in the recent years. In general,
it has been found, as noted above, that diversity can in fact be an important
parameter in controlling the dynamics.  In particular, it has been shown
that both excitable and bistable systems can improve their response to an
external stimulus if there is an adequate degree of diversity in the
constituent units (\cite{Tessone2006}).

In this paper we continue our study of the effect of diversity, investigating in
some depth its role in the FHN model of excitable systems. We consider in detail a system of many neurons
coupled either chemically or electrically. We show that in both cases a right
amount of diversity can indeed enhance the response to a periodic external
stimulus and we discuss in detail the difference between the two types of
coupling. The outline of the paper is as follows:
In section \ref{sec:model} we present the FHN model and the coupling schemes
considered. Next, in section \ref{sec:results} we present the main results we
have obtained from numerical simulations. Afterwards, in section \ref{sec:OPE} we develop an approximate theoretical treatment based on an order
parameter expansion which allows us to obtain a quantitative description of the behavior of the model and compare
its predictions with the numerical results.  Finally, in section \ref{sec:conclusions} we summarize the conclusions.
\section{FitzHugh-Nagumo model}
\label{sec:model}
\subsection{Dynamical equations}
Let us consider a system of excitable neurons described by the FHN model.
The dynamical equations describing the activity of a single neuron are:
\begin{eqnarray}
\epsilon \frac{d x}{dt}&=&x(1-x)(x-b)-y+d,
   \label{FHNvoltage-eq}\\
\frac{d y}{dt}&=&x-cy+a,
\label{FHNrecover-eq}
\end{eqnarray}
where $x(t)$ and $y(t)$ represent, respectively,  the fast membrane potential
and the slow potassium gating variable
of a neuron. We assume that the time scales of these variables are well
separated by the small parameter $\epsilon=0.01$. Other parameters are fixed to
$b=0.5$, $c=4.6$ and $d=0.1$ (\cite{Glatt08}), while the value of $a$ will
fluctuate from one neuron to the other, so reflecting the intrinsic diversity in
the neuronal ensemble.

\begin{figure}[ht!]
 \centering
 \includegraphics[width=0.495\linewidth,clip]{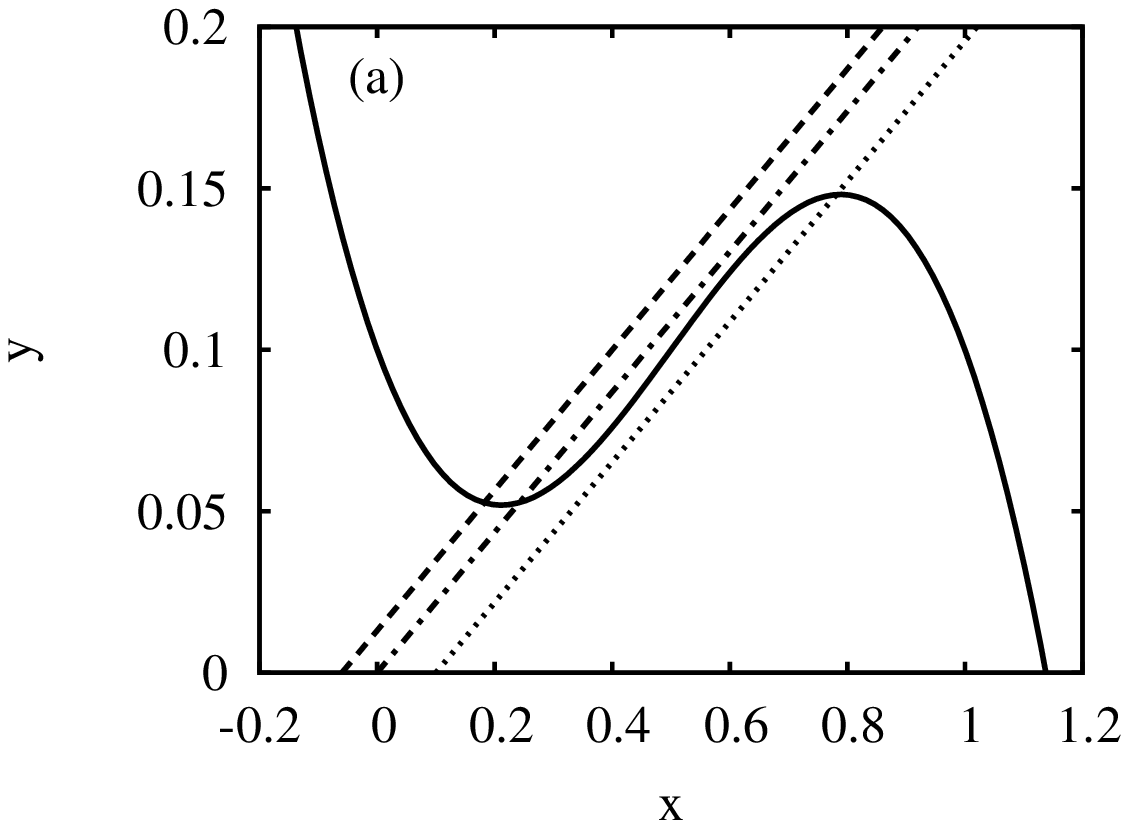}
 \includegraphics[width=0.495\linewidth,clip]{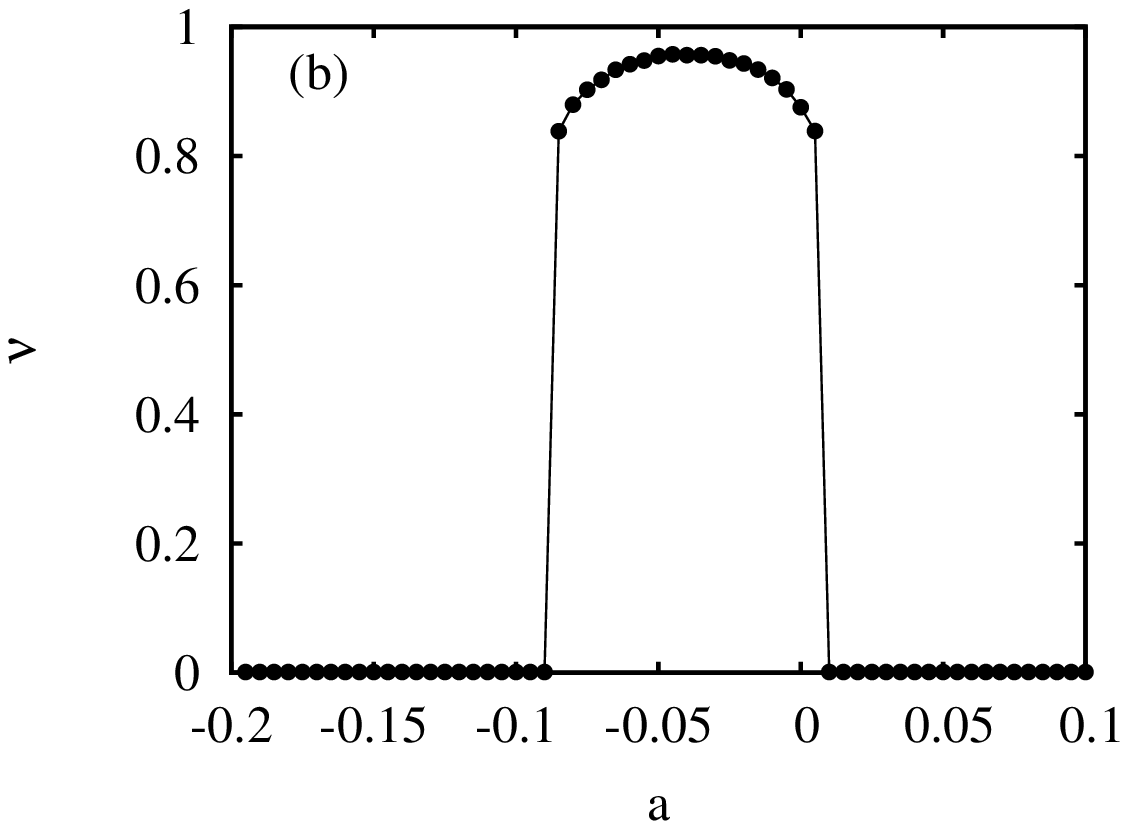}
 \caption{(a) Nullclines of the FHN system for three different values of the
parameter $a$.
$f(x)$ (solid line) and $g(x,a)$ for $a=-0.1$ (dotted line); $a=0.0$
(dash-dotted line) and
$a=0.06$ (dashed line). (b)
Dependence of the firing frequency $\nu$ on $a$.}
 \label{f0FHN}
\end{figure}

\begin{figure}[htb]
\centering
\includegraphics[width=0.99\linewidth,height=4cm,clip]{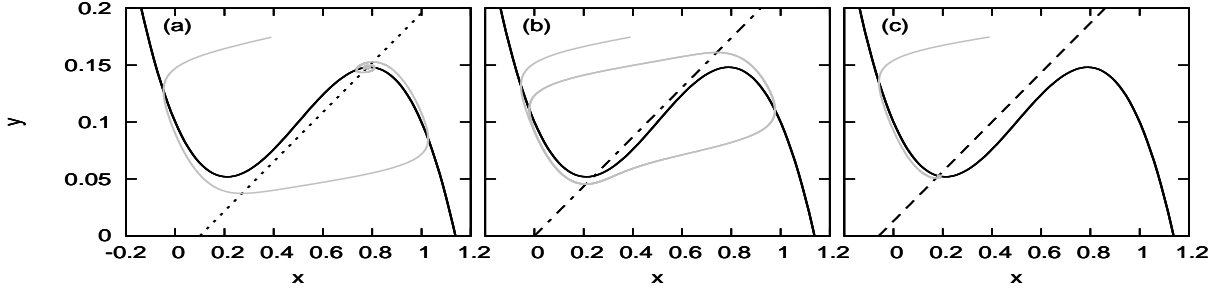}
\caption{Phase-space portrait of the FHN system for different values of $a$.
Gray line represents
the evolution of $\{x,y\}$. $f(x)$ (black line) and $g(x,a)$ for (a)
$a=-0.1$ dotted line (excitable regime); (b) $a=0.0$ dash-dotted line
(oscillatory regime) and
(c) $a=0.06$ dash line (excitable regime).}
\label{f2}
\end{figure}

Let us concentrate first on the dynamics of a single neuron as described by Eqs.
(\ref{FHNvoltage-eq})-(\ref{FHNrecover-eq}). This dynamics has a
strong dependence on the parameter $a$. Three different operating
regimes are identify: for $a\lesssim-0.09$ the system has a stable focus in the
right branch of the cubic nullcline leading the system to an excitable regime;
for $-0.09\lesssim a\lesssim0.01$ a limit cycle around an unstable focus
appears (oscillatory regime) and for $0.01\lesssim a$ a stable focus appears
again, now at the left side of the cubic nullcline (excitable regime). Figure
\ref{f0FHN} (a) shows the nullclines $f(x)=x(1-x)(x-b)+d$ and $g(x,a)=(x+a)/c$
of the system in the three operating regimes described above, for $a=-0.1$, $0.0$ and $0.06$, respectively. In
the excitable regime, spikes (also known as pulses), can appear as a result of
an external perturbation of large enough amplitude. A convenient definition is
that a spike appears when the membrane potential
of the neuron exceeds a certain threshold value, e.g. $x\geq 0.5$. In the
oscillatory regime, spikes appear spontaneously with an intrinsic firing
frequency $\nu$ which, as shown in Figure \ref{f0FHN} (b), does not depend much
on the value of $a$.
\\
To illustrate the dynamical behavior of $x(t)$ and $y(t) $, we show in Figure
\ref{f2} the phase-portrait for three different values of $a=-0.1$, $0.0$ and $0.06$
corresponding to the three nullclines represented in Figure \ref{f0FHN} (a).

\subsection{Coupling scenarios: electrical versus chemical interaction}
\label{diffcoup}

We consider now that we have an ensemble of $N$ coupled neurons. Each one is
described by dynamical variables $x_i(t),\, y_i(t)$, $i=1,\dots,N$, obeying the
FitzHugh-Nagumo equations. The neurons are not isolated from each other, but
interact mutually. The full set of equations is now:
\begin{eqnarray}
\epsilon \frac{d x_i}{dt}&=&x_i(1-x_i)(x_i-b)-y_i+d+I^{syn}_i(t),
   \label{FHNvoltage-eqi1}\\
\frac{d y_i}{dt}&=&x_i-cy_i+a_i,
\label{FHNrecover-eqi}
\end{eqnarray}
where $I^{syn}_i(t)$ is the coupling term accounting for all the interactions of
neuron $i$ with other neurons.
To take into account the natural diversity of the units, we assume that
the parameter $a_i$, that controls the degree of excitability of the neuron,
varies from neuron to neuron. In particular we assign to $a_i$ random values
following a Gaussian distribution with mean
$\left\langle a_i \right\rangle= a$ and correlations
$\left\langle (a_i- a)(a_j- a) \right\rangle=\delta_{ij}\sigma^2$. We use
$\sigma$ as a measure of the heterogeneity of the system and, in the following,
we use the value of $\sigma$ as an indicator of the degree of diversity. If
$\sigma=0$, all neurons have exactly the same set of parameters, while large
values of $\sigma$ indicate a large dispersion in the dynamical properties of
individual neurons.

The most common way of communication between neurons is via chemical synapses,
where the transmission is carried by an agent called neurotransmitter. In these
synapses, neurons are separated by a synaptic cleft and the neurotransmitter has
to diffuse to reach the post-synaptic receptors. In the chemical coupling case
the synaptic current is modeled as:
\begin{eqnarray}
I^{syn}_i(t)=\frac{K}{N_c}\sum_{j=1}^{N_c} g_{ij}r_j(t)\left(E_i^s-x_i\right).
\label{eqnIsyn}
\end{eqnarray}
In this configuration, we consider that neuron $i$ is connected to $N_c$ neurons
randomly chosen from the set of $N-1$ available neurons. Once a connection
is established between neuron \textit{i} and \textit{j}, we assume that the reciprocal
connection is also created. Then, the connection fraction of each neuron is defined as $f=N_c/(N-1)$. 
In Eq. (\ref{eqnIsyn}) K determines the coupling strength and $g_{ij}$ represents the maximum conductance 
in the synapse between the neurons \textit{i} and \textit{j}. For simplicity, we
limit our study to the homogeneous coupling configuration, where $g_{ij}=1$ if
neurons $i$ and $j$ are connected and $g_{ij}=0$ otherwise.  
The character of the synapse is determined by the synaptic reversal potential of the receptor neuron,
$E_i^s$. An excitatory (resp. inhibitory) synapse is characterized by a value of
$E_i^s$ greater (resp. smaller) than the
membrane resting potential. We consider $E_i^s=0.7$ for the
excitatory synapses and $E_i^s=-2.0$ for the inhibitory ones. We also define the fraction
of excitatory neurons (those that project excitatory synapses) in the system as $f_e=N_e/N$ being $N_e$ 
the number of excitatory neurons.

 Finally, $r_j(t)$ is a time dependent function
representing the fraction of bound receptors and it is given by:
\begin{equation}
r_j(t)=\cases{
1-e^{-\alpha t} & {\rm for}\  $t\leq t_{on}$,\cr
 \left(1-e^{-\alpha t_{on}}\right)e^{-\beta \left( t-t_{on}\right)} & {\rm for}\, $t > t_{on}$
}
\end{equation}
where $\alpha=2.5$ and $\beta=3.5$ are the rise and decay time constants,
respectively.
Here $t_{on}=0.1$ represents the time the synaptic connection remains active and $t$
is the time from the spike generation in the presynaptic neuron.

There is another type of synapse where the membranes of the neurons are in close contact and thus the
transmission of the signal is achieved directly (electrical synapses). In this case of
electrically-mediated interactions, also known as diffusive coupling, the total synaptic current is
proportional to the sum of the membrane potential difference between a neuron
and its neighbors, and it is given by:
\begin{eqnarray}
I^{syn}_i(t)=\frac{K}{N_c}\sum_{j=1}^{N_c} \left( x_j-x_i\right).
\end{eqnarray}%

The last ingredient of our model is the presence of an external forcing that
acts upon all neurons simultaneously. Although our results are very general, for
the sake of concreteness we use a periodic forcing of amplitude $A$ and period
$T$. More precisely, the dynamical equations under the presence of this forcing
are modified as:
\begin{eqnarray}
\epsilon \frac{d x_i}{dt}&=&x_i(1-x_i)(x_i-b)-y_i+d+I^{syn}_i(t),
   \label{FHNvoltage-eqi}\\
\frac{d y_i}{dt}&=&x_i-cy_i+a_i+A\sin\left(\frac{2\pi}{T}t\right),
\label{FHNrecover-eqi2}
\end{eqnarray}
which is the basis of our subsequent analysis.

\section{Results}
\label{sec:results}

We are interested in analysing the response of the global system to the external
forcing. We will show that its effect can be enhanced under the presence of the
right amount of diversity in the set of parameters $a_i$, i.e. a conveniently
defined response will reach its maximum amplitude at an intermediate value of the root
mean square value $\sigma$.

In order to quantify the global response of the system with respect to the diversity, we use
the spectral amplification factor defined as
\begin{eqnarray}
\eta &=& \frac{4}{A^{2}} \left| \left\langle e^{-i\frac{2\pi}{T}t} X(t)
\right\rangle\right|^2. \label{saf}
\end{eqnarray}
where $X(t)=\frac{1}{N} \sum_{i=1}^N x_i(t)$ is the global, average collective
variable of the system and $\left\langle \cdot \right\rangle$ denotes a time average. We will analyze
separately the cases of electrical and chemical coupling.
\subsubsection{Electrical coupling}
\begin{figure}[h]
 \centering
 \includegraphics[width=0.99\linewidth,clip]{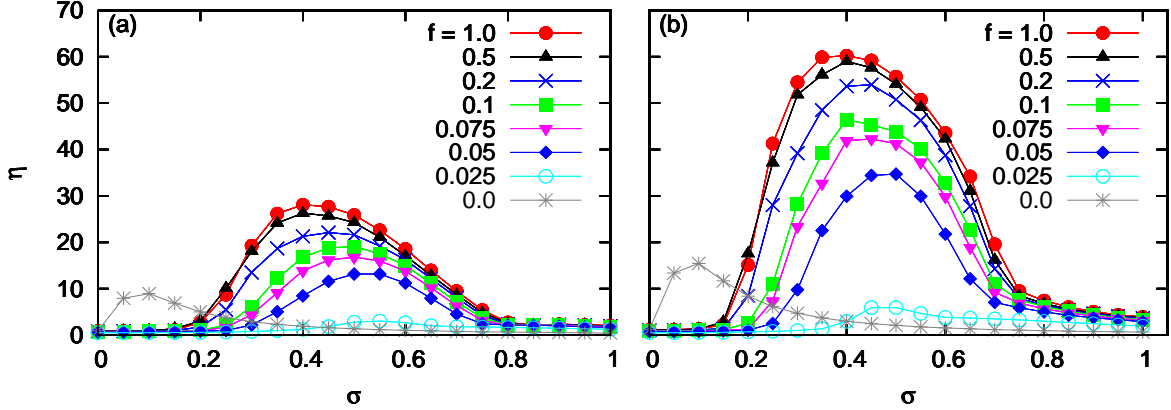}
 \caption{Spectral amplification factor $\eta$ as a function of $\sigma$ for an
increasing fraction of connected neurons $f$ for two different periods of the
external modulation.
(a) $T=1.6$ and (b) $T=1.11$. Other parameters: $a =0.06$, $K=0.6$ and
$A=0.05$.}
 \label{f1FHN}
\end{figure}

\begin{figure}[ht!]
 \centering
 \includegraphics[width=0.95\linewidth,clip]{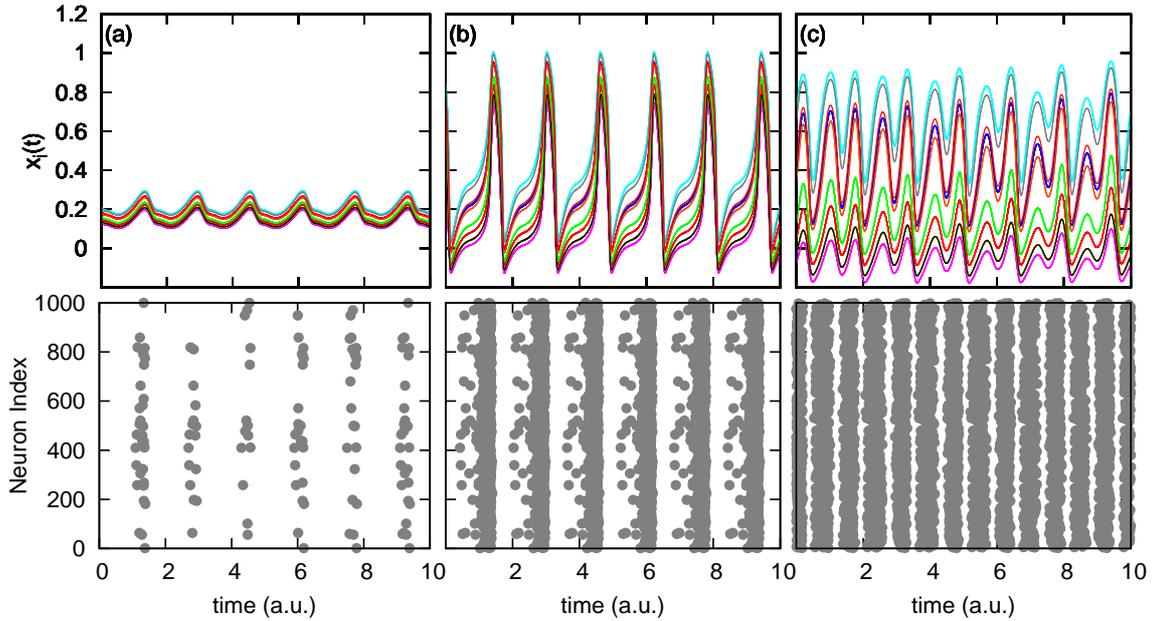}
 \caption{Time traces of ten randomly chosen neurons and raster plot (every time a neuron spikes a dot is drawn) of the
fully connected, $f=1$, ensemble in the case of electrical coupling for three
different values of the diversity parameter: (a)
$\sigma=0.1$, (b) $\sigma=0.4$ and (c) $\sigma=0.9$. Other parameters: $a
=0.06$, $K=0.6$, $A=0.05$ and $T=1.6$.}
 \label{f1bFHN}
\end{figure}

In this subsection we concentrate on the situation in which the neurons are
electrically (diffusively) coupled in a random network, where a neuron \textit{i} is connected
randomly with $N_c=f(N-1)$ other neurons. The mean value of the Gaussian
distribution of the parameters $a_i$ is fixed to
$a =0.06$ and the coupling strength to $K=0.6$. Figure \ref{f1FHN} shows the
spectral amplification factor $\eta$ as a function of the diversity $\sigma$ for
fixed values of the amplitude $A=0.05$ and two different values of the period $T$ of the external
forcing,  for an increasing connection fraction $f$. It can be seen from
Figure \ref{f1FHN}
that intermediate values of $\sigma$ give a maximum response in the spectral
amplification factor.
Moreover, the maximum value shifts slightly to smaller values of $\sigma$ as the
fraction of
connected neurons $f$ increases. We have also observed that a period $T$ of the
external forcing close
to the inverse of the intrinsic firing frequency of the neurons
($\nu\approx0.9$, according to Figure \ref{f0FHN}b) yields the largest response.
\\
In order to further illustrate the response of the system to the external
sinusoidal
modulation, Figure \ref{f1bFHN} shows the raster plot of the ensemble (lower
panels) and the
time traces of ten randomly chosen neurons (upper panels) for different values
of the diversity
parameter.  It can be seen in both the top and bottom panels of this figure that
an intermediate level
of diversity gives a more regular behavior than either smaller or larger values
of $\sigma$.  This fact is more
evident in the time traces where the amplitude of the oscillation varies
randomly for large $\sigma$ values.

\subsubsection{Chemical coupling}
\begin{figure}[h]
 \centering
 \includegraphics[width=0.99\linewidth,clip]{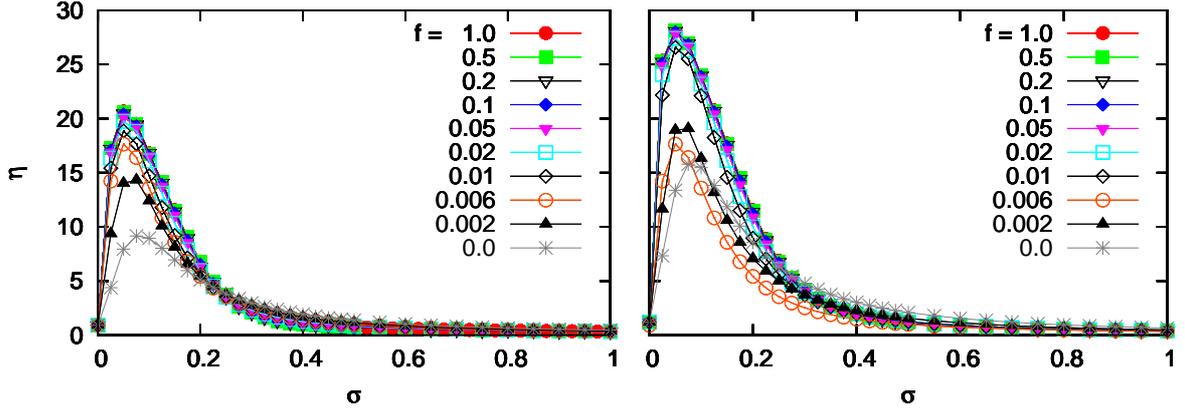}
 \caption{Spectral amplification factor $\eta$ as a function of $\sigma$ for an
increasing fraction of connected neurons $f$. The fraction of excitatory neurons 
was fixed to $f_e=0.8$. Two different periods of the external modulation have been
considered: (a) $T=1.6$ and (b) $T=1.11$.}
 \label{f2FHN}
\end{figure}

\begin{figure}[h]
 \centering
 \includegraphics[width=0.6\linewidth,clip]{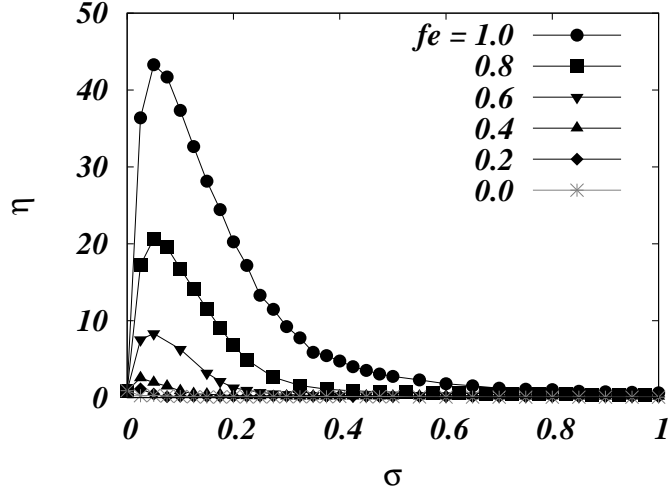}
 \caption{Spectral amplification factor $\eta$ as a function of $\sigma$ for an
increasing fraction of excitatory synapses $f_e$ in the case that neurons are globally coupled, $f=1$. Other parameters: $a=0.06$,
$K=1.5$, $A=0.05$ and $T=1.6$.}
 \label{f3FHN}
\end{figure}

We consider in this subsection the situation in which the units are chemically
coupled.  Figure
\ref{f2FHN} shows the spectral amplification factor $\eta$ as a function of the
diversity $\sigma$
for fixed values of the amplitude $A=0.05$ and two different periods of the
external
modulation when the fraction of randomly connected neurons $f$ increases.
The coupling strength is fixed to $K=1.5$. The fraction of excitatory neurons in the system 
is set to $f_e=0.8$.
It can be seen from Figure \ref{f2FHN} that the spectral
amplification factor $\eta$ increases as $f$ increases, reaching the maximum
response at $f \simeq 0.05$. Interestingly, beyond
this value $\eta$ does not change significantly, indicating that the response of
the system does not improve when the percentage of connected
neurons is larger than $5\%$.  Or, put in another way, with a $5\%$
connectivity, the system already behaves as being fully
connected as far as the response to the external forcing is concerned.  It is also
worth noting that the maximum response is always
at the same value of $\sigma$, independent of $f$.
The effect of changing the ratio of
excitatory/inhibitory synapses in our system is shown in Figure \ref{f3FHN} in the globally coupled case $f=1$. The
spectral amplification factor increases as the fraction of excitatory
connections $f_e$
increases, while the position of the maximum shifts slightly to larger values of
$\sigma \simeq 0.05$.

Comparing the results from both electrical and chemical coupling schemes, it can
be seen that the
electrical coupling gives a larger value of $\eta$ but requires, at the same
time, a larger diversity.
The electrical coupling also exhibits a
larger range of diversity values for which the system has an optimal response
compared with the chemical
coupling. In contrast, the optimal response in the
chemical coupling scheme occurs for small values of the diversity and does not
significantly change in amplitude
and width when the percentage of connected neurons $f$ is increased above $5\%$.
\section{Order Parameter Expansion}
\label{sec:OPE}
It is possible to perform an approximate analysis of the effect of the diversity
in the case of  diffusive (electrical) coupling. The analysis allows to gain
insight into the amplification mechanism by showing how the effective nullclines
of the global variable $X(t)$ are modified when varying $\sigma$. The
theoretical analysis is based upon a modification of the so-called order
parameter expansion developed by Monte and D\textquoteright Ovidio
(\cite{Monte03a,Monte05}) along the lines  of \cite{KT:2010}. The approximation
begins by expanding the
dynamical variables around their average values $X(t)=\frac{1}{N}\sum_i x_i$
and
$Y(t)=\frac{1}{N}\sum_i y_i$ as $x_j(t)=X(t)+\delta^x_j(t)$,
$y_j(t)=Y(t)+\delta^y_j(t)$ and the diversity
parameter around its mean value $a_j=a+\delta^a_j$. The validity of this
expansion relies on
the existence of a coherent behavior by which the individual units $x_j$ deviate
in a small amount $\delta^x_j$ from the global behavior characterized by the
global average variable $X(t)$. It also assumes that the deviations $\delta^a_j$
are small. We expand equations (\ref{FHNvoltage-eqi})
 for $\frac{dx_i}{dt}$ and (\ref{FHNrecover-eqi2}) for $\frac{dy_i}{dt}$
up to second order in $\delta^x_i$, $\delta^y_i$ and $\delta^a_i$; the resulting
equations are:

\begin{eqnarray}
\epsilon\frac{dx_i}{dt}&=&f\left(X,Y\right)+f_x\left(X,Y\right)\delta^x_i+
f_y\left(X,Y\right)\delta^y_i+\frac{1}{2}f_{xx}\left(X,Y\right)
\left(\delta^x_i\right)^2  , \label{exp1-eq}\\
\frac{dy_i}{dt}&=&g\left(X,Y,a\right)+g_x\left(X,Y,a\right)\delta^x_i+
g_a\left(X,Y,a\right)\delta^a_i , \label{exp2-eq}
\end{eqnarray}
where
\begin{equation}
\begin{array}{rcl}
f(x,y)&=& x(1-x)(x-b)-y+d-Kx , \cr
g(x,y,a)&=&x-cy+a+A\sin\left(\frac{2\pi}{T} t\right),
\end{array}
\end{equation}
and we used the notation $f_x$ to indicate the derivative of $f$ with respect to
$x$ and so forth.
Note that Eq. (\ref{exp2-eq}) is exact since it is linear in all the variables.
If we average Eq. (\ref{exp1-eq})-(\ref{exp2-eq}) using $\langle \cdot
\rangle=\frac{1}{N}\sum_i\cdot$ we obtain:
\begin{eqnarray}
\epsilon\frac{dX}{dt}&=&f\left(X,Y\right)+
\frac{1}{2}f_{xx}\left(X,Y\right)\Omega^x ,\label{exp3-eq}\\
\frac{dY}{dt}&=& g\left(X,Y,a\right) , \label{exp4-eq}
\end{eqnarray}
where we have used $\langle \delta^x_j \rangle=\langle
\delta^y_j \rangle=\langle \delta^a_j \rangle=0$ and defined the
second moment $\Omega^x=\langle(\delta^x_j)^2 \rangle$. We also defined
$\Omega^y=\langle(\delta^y_j)^2 \rangle$,
$\sigma^2=\langle(\delta^a_j)^2 \rangle$, and the shape factors
$\Sigma^{xy}=\langle\delta^x_j\delta^y_j\rangle$,
$\Sigma^{xa}=\langle\delta^x_j\delta^a_j\rangle$ and
$\Sigma^{ya}=\langle\delta^y_j\delta^a_j\rangle$.
The evolution equations for the second moments are found from
the first-order expansion $\dot{\delta}^x_j=\dot{x}_j-\dot{X}$, so
that $\dot{\Omega}^x=2\langle\delta^x_j\dot{\delta}^x_j\rangle$
and
$\dot{\Sigma}^{xy}=\langle\dot{\delta}^x_j\delta^y_j+\delta^x_j\dot{
\delta}^y_j\rangle$, were the dot stands for time derivative.
\begin{eqnarray}
\epsilon\dot{\delta}^x_i&=&f_x\delta^x_i+f_y\delta^y_i+\frac{1}{2}f_{xx}
\left[\left(\delta^x_i\right)^2-
\Omega^x\right] ,\label{exp5-eq} \\
\dot{\delta}^y_i&=& g_x\delta^x_i+g_y\delta^y_i+g_a\delta^a_i ,
\label{exp6-eq}\\
\dot{\Omega}^x&=&\frac{2}{\epsilon}\left[f_x\Omega^x+
f_y\Sigma^{xy}\right] ,\label{exp7-eq}\\
\dot{\Omega}^y&=&2\left[g_x\Sigma^{xy}+g_y\Omega^{y}+g_a\Sigma^{ya}\right] ,
\label{exp8-eq}\\
\dot{\Sigma}^{xy}&=&\frac{1}{\epsilon}\left[f_x\Sigma^{xy}+
f_y\Omega^y\right]+g_x\Omega^x+g_y\Sigma^{xy}+g_a\Sigma^{xa} ,\label{exp9-eq}\\
\dot{\Sigma}^{xa}&=&\frac{1}{\epsilon}\left[f_x\Sigma^{xa}+
f_y\Sigma^{ya}\right] ,\label{exp10-eq}\\
\dot{\Sigma}^{ya}&=&g_x\Sigma^{xa}+g_y\Sigma^{ya}+g_a\sigma^2 .\label{exp11-eq}
\end{eqnarray}

The system of Eq. (\ref{exp3-eq})-(\ref{exp4-eq}) together with Eq.
(\ref{exp7-eq})-(\ref{exp11-eq}) forms a closed set of differential equations
for the average collective variables $X(t)$ and $Y(t)$:

\begin{eqnarray}
\epsilon\dot{X}&=&-X^3+(1+b)X^2-(b+3\Omega^x)X+(1+b)\Omega^x+d-Y ,
\label{exp12-eq}\\
\dot{Y}&=& X-cY+a+A\sin\left(\frac{2\pi}{T} t\right) ,\label{exp13-eq}\\
\epsilon\dot{\Omega}^x &=& 2(-3X^2+2(1+b)X-b-K)\Omega^x-2\Sigma^{xy} ,
\label{exp14-eq}\\
\dot{\Omega}^y&=&2\left[\Sigma^{xy}-c\Omega^y+\Sigma^{ya}\right] ,
\label{exp15-eq}\\
\dot{\Sigma}^{xy}&=&\frac{1}{\epsilon}\left[(-3X^2+2(1+b)X-b-K)\Sigma^{xy}
-\Omega^y\right]{} \nonumber \\ & & {}
+\Omega^x-c\Sigma^{xy}+\Sigma^{xa} ,\label{exp16-eq}\\
\epsilon\dot{\Sigma}^{xa}&=&(-3X^2+2(1+b)X-b-K)\Sigma^{xa}
-\Sigma^{ya} ,\label{exp17-eq}\\
\dot{\Sigma}^{ya}&=&\Sigma^{xa}-c\Sigma^{ya}+\sigma^2 .\label{exp18-eq}
\end{eqnarray}
Numerical integration of this system allows us to obtain $X(t)$, from which
we can compute the spectral amplification factor $\eta$. The value of $\eta$
obtained
from the expansion is later compared with that obtained from the numerical
integration of
the Eqs. (\ref{FHNvoltage-eqi})-(\ref{FHNrecover-eqi}) (see Figure \ref{f1theo}
below).

\begin{figure}[h!]
\centering
\includegraphics[width=0.65\linewidth,clip]{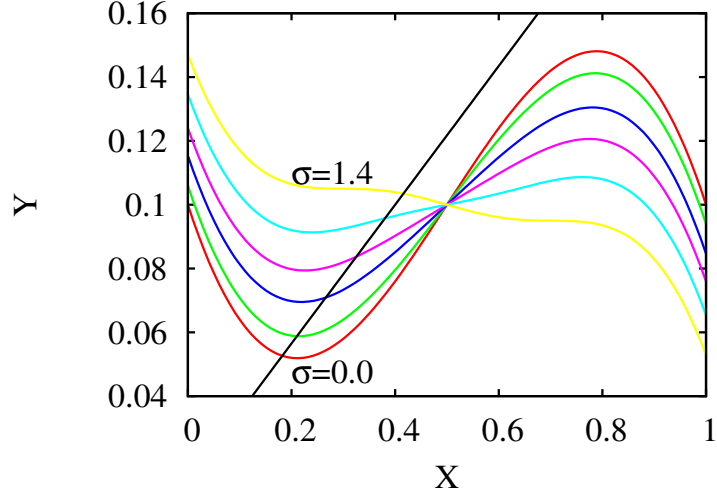}
\caption{Nullclines of Eq. (\ref{MF1})-(\ref{MF2}) for different values of the
diversity $\sigma$. $Y_1(X,\sigma)$ for
$\sigma$: (a) $0.0$ (red line), (b) $0.5$ (green line), (c) $0.8$ (blue line), (d) $1.0$ (violet line), 
(e) $1.2$ (cian line) and (f) $1.4$ (yellow line). The nullcline $Y_2(X,a)$ of Eq. (\ref{MF2}) 
for $a=0.06$ is represented with a black line.}
\label{fig:f03}
\end{figure}

\begin{figure}[h!]
\centering
\includegraphics[width=0.99\linewidth,clip]{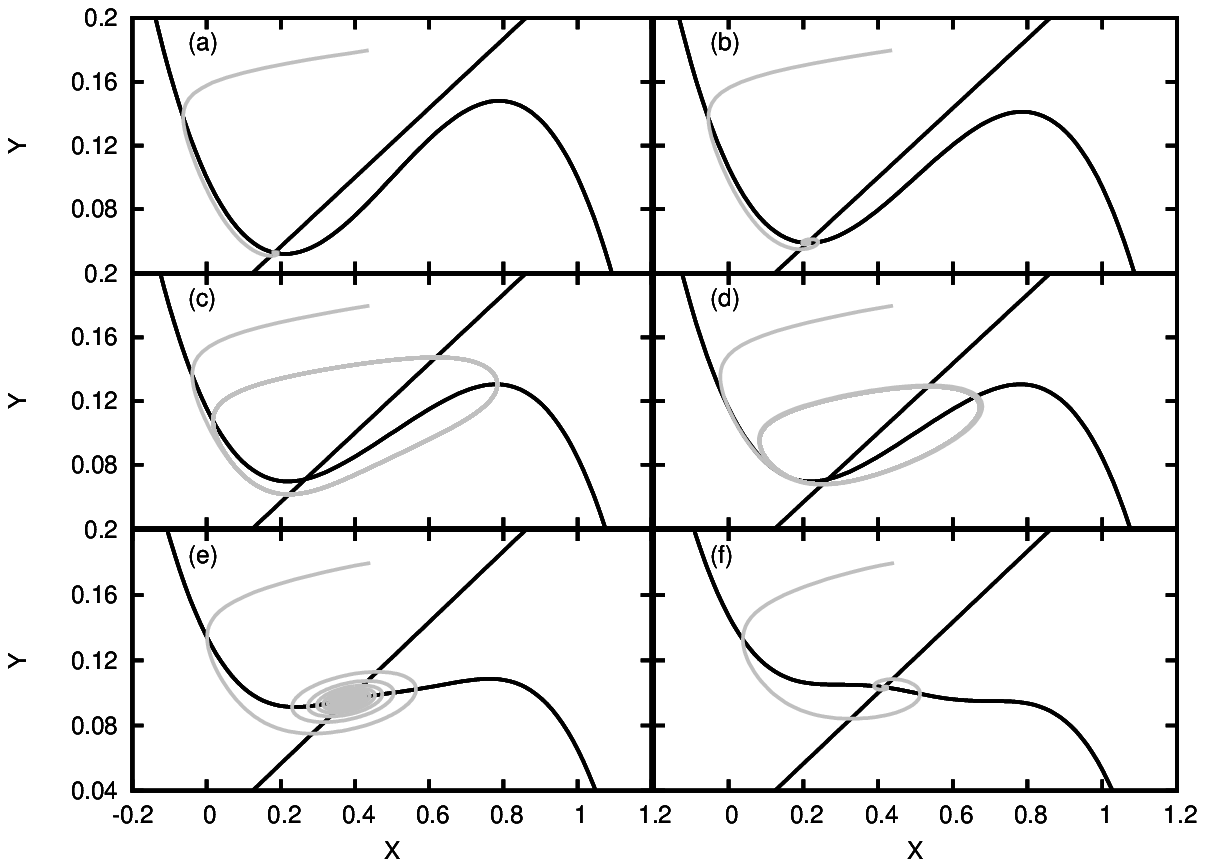}
\caption{Phase-space portrait for different values of $\sigma$. Grey line
represents the evolution
of $\{X,Y\}$. The black lines represent the nullcline $Y_2(X,a)$ for $a=0.06$ 
and the cubic nullcline $Y_1(X,\sigma)$ for
$\sigma$: (a) $0.0$, (b) $0.5$, (c) $0.8$, (d) $1.0$, (e) $1.2$ and (f)
$1.4$.}
\label{fig:f04}
\end{figure}

We can obtain another set of closed equations for $X(t)$ and $Y(t)$ if we
perform an adiabatic
elimination of the fluctuations, i.e.,
$\dot{\Omega}^x=\dot{\Omega}^y=\dot{\Sigma}^{xy}=\dot{\Sigma}^{xa}=\dot{\Sigma}^
{ya}=0$, yielding to:
\begin{equation}
\begin{array}{lll}
\displaystyle \Sigma^{xa}=\frac{\sigma^2}{cH(x)-1} , & \Sigma^{ya}=\frac{H(x)\sigma^2}{cH(x)-1}, &\Sigma^{xy}=\frac{H(x)\sigma^2}{(cH(x)-1)^2} ,\label{adiabatic_3}\cr
\displaystyle \Omega^x=\frac{\sigma^2}{(cH(x)-1)^2} ,& \Omega^y=\frac{H^2(x)\sigma^2}{(cH(x)-1)^2} ,
 \end{array}
 \end{equation}
with $H(x)=-3x^2+2(1+b)x-b-K$.
Substituting $\Omega^x$ in Eqs. (\ref{exp12-eq})-(\ref{exp13-eq}), we find
a
closed form for the equations describing the evolution of the mean-field
variables $X(t)$
and $Y(t)$:
  \begin{eqnarray}
 \epsilon\dot{X}&=&-X^3+(1+b)X^2-\left[b+\frac{3\sigma^2}{(cH(X)-1)^2}\right]X
+\frac{(1+b)\sigma^2}{(cH(X)-1)^2}+d-Y,  \label{MF1}\\
 \dot{Y}&=&X-cY+a +A\sin\left(\frac{2\pi}{T} t\right) \label{MF2}
 \end{eqnarray}
These equations provide a closed form for the nullclines of
the global
variables $X$ and $Y$ for the non-forcing case $A=0$. They also reflect how diversity influences these
variables.
Figure \ref{fig:f03} shows these nullclines $Y_1(X,\sigma)$ and $Y_2(X,a)$ of
Eqs. (\ref{MF1})-(\ref{MF2}) respectively for
$a=0.06$ and different values of the diversity $\sigma$.
It can be seen in the figure
that the diversity changes the shape of the cubic nullcline
$Y_1(X,\sigma)$ leading to a loss of stability of the fixed point of the system
that, for a
certain range of $\sigma$, becomes a limit cycle.
To schematize the behavior of the global variables $X$ and $Y$ when the
diversity changes, we show in Figure \ref{fig:f04} the phase-portrait for
different values of $\sigma=0.0$, $0.5$, $0.8$, $1.0$, $1.2$ and $1.4$
(corresponding to the values represented in Figure \ref{fig:f03}). It can be
seen that there is a range of $\sigma$ for which the system exhibits a
collective oscillatory behavior even in the absence of the weak external
modulation. \\
With the collective variable $X(t)$ obtained from the adiabatic elimination we
can estimate the spectral amplification factor $\eta$.  Figure \ref{f1theo} shows the results obtained from the numerical integration
of Eqs. (\ref{exp12-eq})-(\ref{exp18-eq}),
together with the numerical simulation of the full system, Eqs.
(\ref{FHNvoltage-eqi})-(\ref{FHNrecover-eqi}) and the adiabatic
approximation obtained from Eqs. (\ref{MF1}) and (\ref{MF2}). It can be seen that our order parameter expansion is in good agreement
with the numerical integration of the full system, even in the case in which the
second
moments are adiabatically eliminated.
\begin{figure}[h!]
 \centering
 \includegraphics[width=0.99\linewidth,clip]{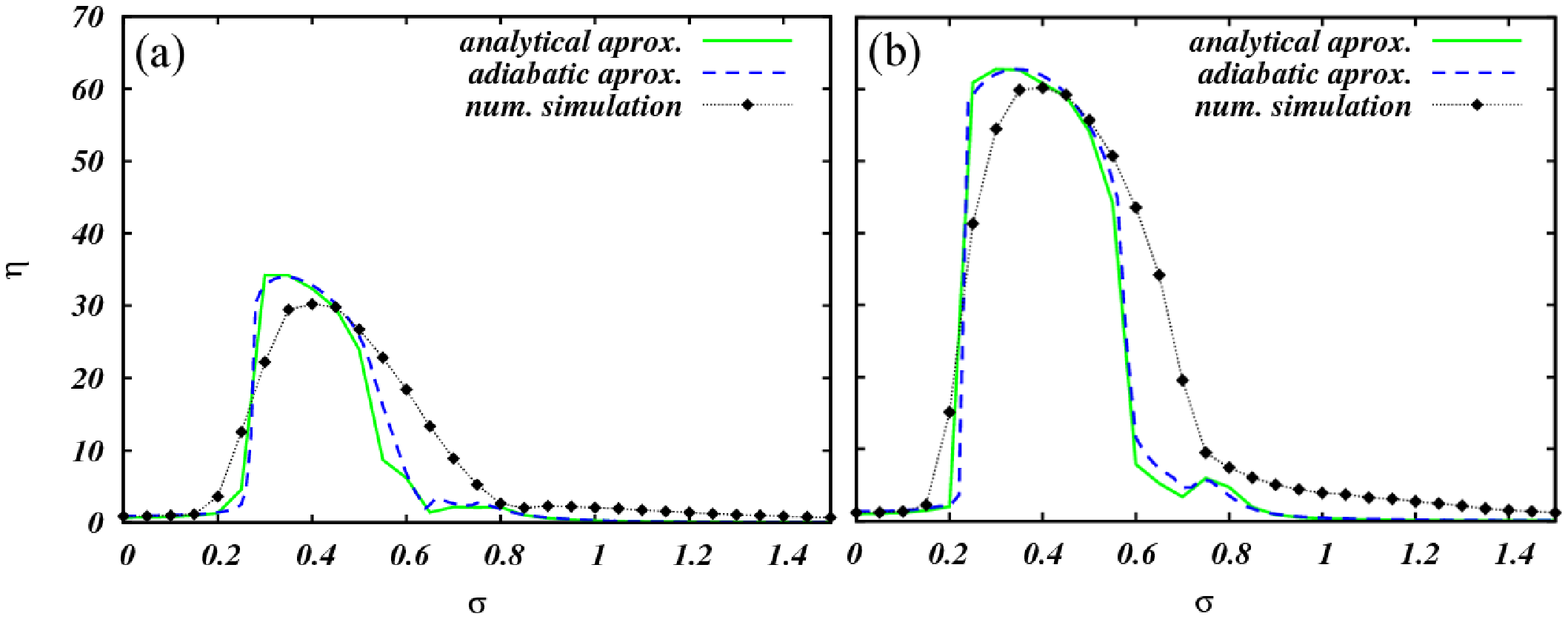}
 \caption{Order parameter expansion versus numerical integration of the
full system. An adiabatic approximation is also included (see text). Two
different periods of the
external modulation have been considered: (a) $T=1.6$ and (b) $T=1.11$.
Other parameter as in Figure \ref{f1FHN}.}
 \label{f1theo}
\end{figure}

\section{Conclusions}
\label{sec:conclusions}
We have studied the effect of the diversity in an ensemble of coupled
neurons described by the FHN model. We have observed that an intermediate
value of diversity can enhance the response of the system to an
external periodic forcing. We have studied both electrical and
chemical coupling schemes finding that the electrical coupling
induces a larger response of the system
to an external weak modulation, as well as existing for a wider range. In
contrast, the chemical coupling scheme exhibits a smaller optimal amplitude and
narrower range of response, however, for smaller values of the diversity. We have also
found that the response of the system in the electrical
coupling scheme strongly depends on the fraction of connected neurons in
the system whereas it does not improve much above a
small fraction of connected neurons in the chemical coupling scheme.
We have also developed an order parameter
expansion whose results are in good agreement with those obtained numerically for
the electrically (diffusively) coupled FHN system. By an adiabatic elimination of the fluctuations
we have found a closed form of the effective nullclines
of the global collective variables of the system obtaining a simple expression
of how the diversity influences the collective variables of the system.

The microscopic mechanism leading the system to a resonant
behavior with the external signal is as follows: in the homogeneous situation,
where all the units are identical, the weak external modulation cannot induce
any spike in the system. When the diversity increases, a fraction of the neurons
enters into the oscillatory regime and, due to the interactions, pull the other
neurons with them. This leads the system to an oscillatory collective behavior
that follows the
external signal. For larger values of the diversity, the fraction of neurons
in the oscillatory regime decreases and the rest of neurons offer some
resistance to being pulled by the oscillatory ones; thus, the system cannot
respond
to the external signal anymore. These
results suggest that the diversity present in biological systems may have an
important role in  enhancing the response of the system to  the
detection of weak signals.

We acknowledge financial support from the following organizations: 
National Science Foundation (Grant DMR- 0702890);
G. Harold and Leila Y. Mathers Foundation;
European Commission Project GABA (FP6-NEST Contract 043309); EU NoE Biosim (LSHB-CT-2004-005137); and 
MEC (Spain) and Feder (project FIS2007-60327).

\end{document}